\documentclass[a4paper]{jpconf}
\usepackage[dvipdfmx]{graphicx}
\usepackage{physics}
\usepackage{amsmath,amssymb}
\usepackage{ascmac}
\usepackage{physics}
\usepackage{bm}
\usepackage{braket}
\usepackage{slashed}
\usepackage{float}
\usepackage{mathtools}
\usepackage{comment}
\usepackage{color}
\usepackage{amsmath}
\usepackage{cite}

\newcommand{\CP}{\mathbb{C}\mathrm{P}}

\begin{document}

\title{
	$\mathbb{C}\mathrm{P}^2$ Skyrmion with Fermion Backreaction
	}

\author{Yuki Amari$^{1,2}$, Nobuyuki Sawado$^3$ and Shintaro Yamamoto$^{3,}$}

\address{$^1$ Research and Education Center for Natural Sciences, Keio University, Hiyoshi 4-1-1, Yokohama, Kanagawa 223-8521, Japan}
\address{$^2$ Department of Physics, Keio University, 4-1-1 Hiyoshi, Kanagawa 223-8521, Japan}

\address{$^3$ Department of Physics and Astronomy, Faculty of Science and Technology, Tokyo University of Science, Noda, Chiba 278-8510, Japan}

\ead{shintaroyamamoto019@gmail.com}

\begin{abstract}

When fermions interact with a topological soliton, they localize on the soliton. In studies of such systems, solitons are often treated as fixed background fields, and the backreaction due to the fermion localization is usually neglected for simplicity. In this work, we investigate the backreaction of localized fermions on a $\CP^2$ Skyrmion. We find that the Skyrmion profile deforms in response, becoming more concentrated around the localized fermions. We also discuss the possibility that this backreaction may play a role analogous to that of the potential term.
\end{abstract}

\section{Introduction}

In the 1960s, Tony Skyrme proposed a model of nucleons and atomic nuclei using topological solitons, now known as Skyrmions, within a low-energy effective theory of quantum chromodynamics~\cite{Skyrme:1961vq,Skyrme:1962vh}. Since then, Skyrmion-like topological solitons have been found not only in high-energy physics but also in a wide range of physical systems, including magnets~\cite{Mühlbauer:2009,Yu:2010ayi,Nagaosa:2013}, liquid crystals~\cite{2014PhRvE..90a2505A}, and optics~\cite{2018Sci...361..993T,2019NatPh..15..650D,2023LPRv...1700155M,2024NaPho..18...15S}.
In particular, two-dimensional analogues of Skyrmions in magnetic materials have been intensively investigated over the past fifteen years, following their experimental discovery~\cite{Mühlbauer:2009,Yu:2010ayi}. These 2D magnetic Skyrmions are also regarded as promising candidates for applications in nanotechnology.

Since the order parameter space (OPS) of (anti-) ferromagnets is $S^2=\CP^1$, 2D Skyrmions in such a magnet are often referred to as $\CP^1$ Skyrmion. As a natural generalization, 2D Skyrmions with the OPS being $\CP^N=\mathrm{SU}(N+1)/\mathrm{U}(N)$, called  $\CP^N$ Skyrmion, have also been theoretically studied in several condensed matter systems~\cite{2008PhRvL.100d7203I,Garaud:2011zk,Garaud:2012pn,Kovrizhin:2012ff,Akagi:2021dpk,Zhang:2023,Benfenati:2022jvq,Amari:2022boe,2025arXiv250917403L}\footnote{
The manifold $\CP^N$ appears as the OPS of some physical systems. For instance, the OPS of spin-$\frac{l}{2}$ spinor Bose-Einstein Condensates is $\CP^{l}$~\cite{PhysRevB.81.184526}, and that of $M$-layered Hall system is $\CP^{2M-1}$~\cite{PhysRevLett.82.3512}.} . Note that, thanks to the homotopy group $\pi_2(\CP^N)=\mathbb{Z}$ for any positive integer $N$, $\CP^N$ Skyrmions are characterized by an integer topological charge. Since $\CP^N$ Skyrmions possess a large number of degrees of freedom, they reveal exotic pattern formation \cite{Akagi:2021lva,Amari:2022boe} 
and intriguing dynamics~\cite{2025arXiv250917403L}. Such novel properties, absent in $\CP^1$ Skyrmions, may pave new ways in the development of Skyrmion-based nanotechnology. In this article, we study the minimal case, i.e., $\CP^2$ Skyrmions, for simplicity.

It is well known that topological solitons exhibit nontrivial spin–statistics properties determined by their topological charge~\cite{Wilczek:1983,Amari:2019tgs}. 
Like other cases, $\CP^2$ Skyrmions 
support fermionic zero modes~\cite{Amari:2023gjq,Amari:2024rpm}, 
of which number is equal to the topological invariant as shown by 
the Atiyah–Singer index theorem~\cite{Atiyah:1963zz}. 
Occupying these zero modes not only realizes the expected spin–statistics features but also induces a backreaction on the Skyrmion, thereby allowing one to investigate the interplay among fermion number, symmetry constraints, and soliton stability in a dynamical and self-consistent manner.

This article is organized as follows. We begin with a brief introduction to our model describing fermion-$\CP^2$ Skyrmion {interaction} in Sec.\ref{Model}. 
In Sec.\ref{Couple}, we present some typical numerical solutions in the coupled system. Then one finds that fermion {energy} spectra exhibits a characteristic behavior, known as spectral flow, and the bound states are localized around the Skyrmion. We end up with our summary and outlook in Sec.\ref{Summary}.

\section{The Model}
\label{Model}
We consider the $\CP^2$ Skyrmion-Dirac fermion coupled system in $2+1$ dimensions in the form
\begin{align}
	\mathcal{L}
	&=\overline{\Psi}\qty(i\hat{\gamma}^\mu\partial_\mu-m\hat{U})\Psi
	+\frac{M^2}{2}\mathrm{Tr}\qty(\partial_\mu U\partial^\mu U)
	{+e^2\mathrm{Tr}\qty(\qty[\partial_\mu U,\partial_\nu U]^2)}
	-V,\label{FullLagrangian}
\end{align}
with $M^2, e^2>0$. Here, $\Psi$ represents a spinor triplet and $U$ is a Hermitian matrix-valued field defined as $U=I_{3\times 3}-2Z\otimes Z^\dagger$, with the $3\times3$ unit matrix $I_{3\times3}$ and the three-component complex unit vector $Z$. To act the spinor,  $\hat{U}$ and $\hat{\gamma}^\mu$ are defined by $\hat{U}\equiv I_{2\times 2}\otimes U,\ \hat{\gamma}^\mu\equiv \gamma^\mu\otimes I_{3\times 3}$, where $\gamma^\mu$ are the gamma matrices satisfying the Clifford algebra $\qty{\gamma^\mu,\gamma^\nu}=2\eta^{\mu\nu}$ with the Minkowski metric $\eta^{\mu\nu}=\mathrm{diag}\qty(1,-1,-1)$. The two-dimensional version of the gamma matrices are defined by the usual Pauli matrices $\sigma_i~\qty(i=1,2,3)$ as $\gamma^0=\sigma_3$, $\gamma^1=-i\sigma_1$ and $\gamma^2=-i\sigma_2$ in the minimal representation. Each term in the Lagrangian \eqref{FullLagrangian} has the following interpretation: the first term is the Dirac Lagrangian, the second term corresponds to the nonlinear sigma model (NLSM), the third term is the Skyrme term, and the fourth term represents the potential. In this work, we employ the potential term of the form
\begin{equation}
    V=\frac{\mu^2}{16}\qty[\mathrm{Tr}\qty(I_{3\times3}-U_\infty U)]^2
    \label{pot}
\end{equation}
where $\mu^2>0$ and $U_\infty$ is the asymptotic value of the $U$ field.
This potential is the square of the so-called old-baby potential. We employ this because the ordinary old-baby potential does not support $\CP^2$ soliton solutions~\cite{PhysRevD.92.045007}. 

The field $U$ is clearly invariant under the local gauge transformation $Z\to e^{i\theta}Z$ with $\theta\in\mathbb{R}$. Due to the phase redundancy and the nomalization condition $Z^\dagger Z=1$, the field $Z$ takes its value on $S^5/S^1\simeq SU(3)/U(2) \simeq \CP^2$. Since we have a homotopy group $\pi_2\qty(\CP^2)=\mathbb{Z}$, there is a topological index, so called topological charge $Q$. The topological charge $Q$ 
is given by 
\begin{align}
	Q=-\frac{i}{2\pi}\int\dd[2]{x}\epsilon_{jk}\qty(D_jZ)^\dagger D_kZ.
\end{align}
This integer labels the Skyrmion, and corresponds to the number of fermionic zero modes as shown later.

In order to obtain the field equations of the system, we rewrite the Lagrangian \eqref{FullLagrangian} in terms of $Z=\qty(Z_1,Z_2,Z_3)^T$ where $T$ represents transposition of the vector. In this article, we choose the vacuum as $U_\infty=\mathrm{diag}\qty(1,1,-1)$ which corresponds to $Z_\infty\propto\qty(0,0,1)^T$. Then, decomposing the Lagrangian into the fermionic part $\mathcal{L}_f$ and bosonic part $\mathcal{L}_b$, i.e., $\mathcal{L}=\mathcal{L}_f+\mathcal{L}_b$ , one can write them in terms of $Z$ as follows:
\begin{align}
&\begin{aligned}
	\mathcal{L}_f
	=\overline{\Psi}\qty(i\hat{\gamma}^\mu\partial_\mu-mI_{2\times2}\otimes I_{3\times 3})\Psi+2m\overline{\Psi} I_{2\times 2}\otimes\qty(Z\otimes Z^\dagger)\Psi,
\end{aligned}
\\
&\begin{aligned}
	\mathcal{L}_b
	&=4M^2\qty(D_\mu Z)^\dagger D^\mu Z\\
	&~~~+32e^2\qty[2\qty(\qty(D_\mu Z)^\dagger D_\nu Z)^2-\qty(\qty(D_\mu Z)^\dagger D^\mu Z)^2-\abs{\qty(D_\mu Z)^\dagger D_\nu Z}^2]-\mu^2\qty(1-\abs{Z_3}^2)^2,
    \end{aligned}
\end{align}
where $D_\mu\equiv\partial_\mu-Z^\dagger\partial_\mu Z$ is the covariant derivative. Based on the variational principle with the nonlinear constraint $Z^\dagger Z=1$, we obtain the equations of motion of the form
\begin{align}
    \qty(i\gamma^\mu_{\alpha\beta}\delta_{ab}\partial_\mu-m\delta_{\alpha\beta}\delta_{ab}+2m\delta_{\alpha\beta}Z_aZ_b^*)\Psi_{\beta,b}=0\label{Fermioniceq}
\end{align}
\begin{align}
	&-4M^2\qty(D_\mu D^\mu-Z^\dagger D_\mu D^\mu Z)Z_a
	+2\mu^2\qty(1-\abs{Z_3}^2)Z_3\delta_{a,3}\nonumber\\
	&-128e^2\qty{D_\mu\qty{\qty[\qty(D^\mu Z)^\dagger D^\nu Z]D_\nu Z_a}-Z^\dagger D_\mu\qty{\qty[\qty(D^\mu Z)^\dagger D^\nu Z]D_\nu Z} Z_a}\nonumber\\
	&+64e^2D_\mu\qty{\qty[\qty(D_\nu Z)^\dagger D^\nu Z]D^\mu Z_a+\qty[\qty(D^\nu Z)^\dagger D^\mu Z]D_\nu Z_a}\nonumber\\
	&-64e^2Z^\dagger D_\mu\qty{\qty[\qty(D_\nu Z)^\dagger D^\nu Z]D^\mu Z+\qty[\qty(D^\nu Z)^\dagger D^\mu Z]D_\nu Z}Z_a\nonumber\\
	&+2m\qty[\Psi_{\alpha,b}^*Z_b\gamma^0_{\alpha\beta}\Psi_{\beta,a}-\qty(\Psi_{\alpha,b}^*\gamma^0_{\alpha\beta}Z_bZ_c^*\Psi_{\beta,c})Z_a]\nonumber\\
	&=0\label{Bosoniceq}
\end{align}
where $\alpha,\beta=1,2$ denote the spinor indices and $a,b,c=1,2,3$ represent the internal indices.

In this article, we consider a rotationally symmetric static Skyrmion. In order to find the solutions of the field equations \eqref{Fermioniceq} and \eqref{Bosoniceq}, we introduce a convenient parametrization of $Z$. Taking into account the single-valuedness of the field $Z$ and the phase redundancy, such a configuration can be written as 
\begin{align}
	Z
	=\mqty
	(
		\cos F\qty(r),
		\sin F\qty(r)\cos G\qty(r) e^{in_1\varphi},
		\sin F\qty(r)\sin G\qty(r) e^{in_2\varphi}
	)^T.
	\label{Bosonicansatz}
\end{align}
Here $F\qty(r)$ and $G\qty(r)$ are monotonic functions of the radial coordinate $r$, $n_1$ and $n_2$ are integers, and $\varphi$ is the polar coordinate. This ansatz \eqref{Bosonicansatz} automatically satisfies the nonlinear constraint $Z^\dagger Z=1$. In this parametrization, the topological charge $Q$ can be written as
\begin{align}
    Q=\int_0^\infty\dd{r}\partial_r\qty[\sin^2F\qty(n_1\cos^2G+n_2\sin^2G)].
\end{align}
It follows that the topological charge $Q$ depends on the boundary conditions of $F$ and $G$. Since we have $Z_\infty\propto\qty(0,0,1)^T$, they must satisfy $F\qty(\infty)=G\qty(\infty)=\qty(\frac{1}{2}+k)\pi\qty(k\in\mathbb{Z})$. For the regularity of the 
static energy at the origin, $\sin F\qty(0)$ must be $0$, i.e., $F\qty(0)=l\pi\qty(l\in\mathbb{Z})$. 
Therefore, we employ $F\qty(0)=G\qty(0)=0,\ F\qty(\infty)=G\qty(\infty)=\frac{\pi}{2}$. 
Then, the topological charge $Q$ equals to $n_2$. 
In this article, we consider the case with $(n_1,n_2)=(1,3)$, giving $Q=3$, because it is pointed out in Ref.~\cite{PhysRevD.92.045007} that $Q=3$ solutions are the most simple nontrivial one in the Skyrme-type model \eqref{FullLagrangian} with the potential \eqref{pot}. 


\section{Numerical Results}
\label{Couple}
\begin{figure}[t]
	\centering
		\includegraphics[width=0.6\linewidth]{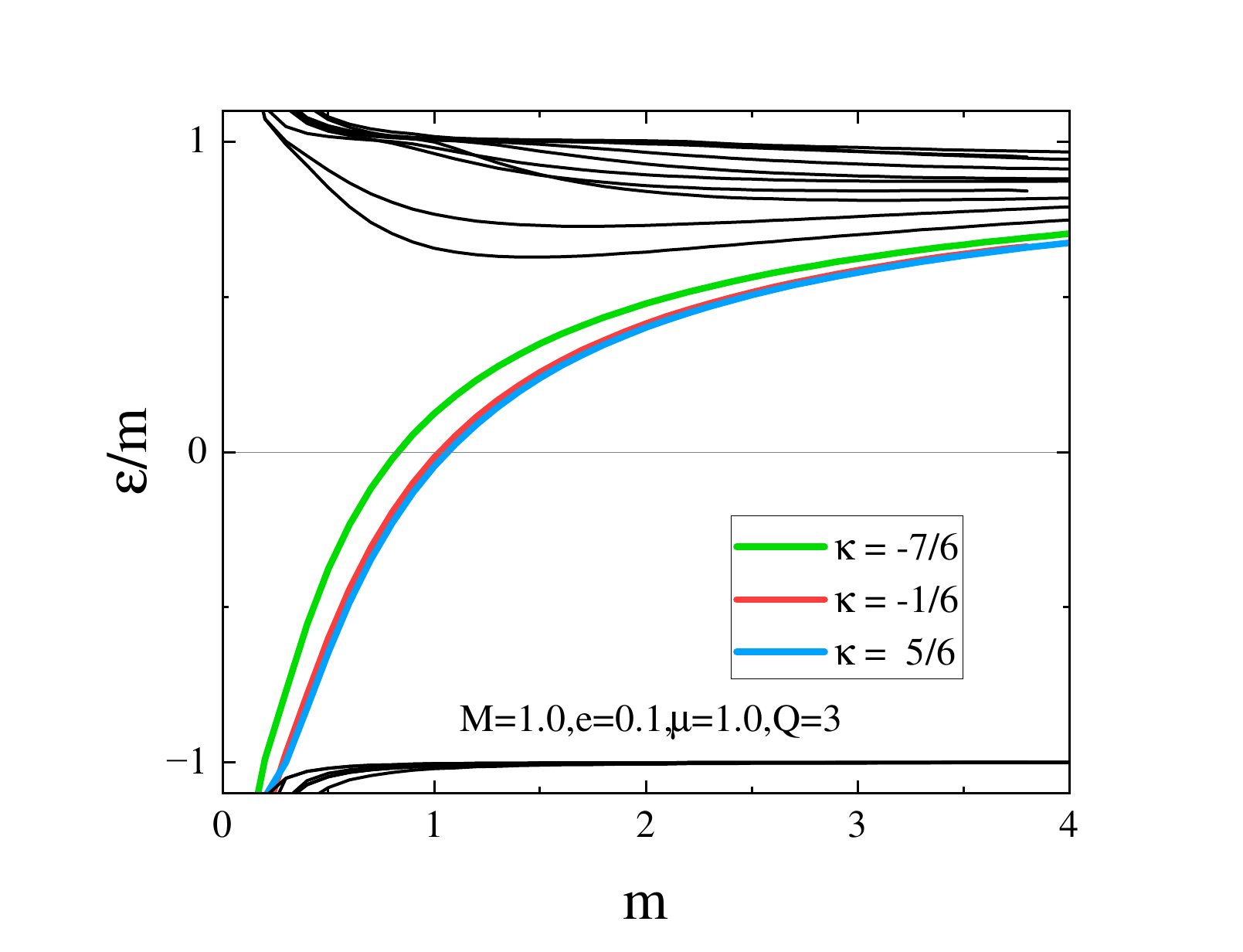}
		\caption{Spectral flow with fermion backreaction. \if0{The parameters in the bosonic sector are fixed to $\qty(M,e,\mu)=\qty(1.0,0.1,1.0)$, and the topological charge $Q=3$.}\fi}
		\label{SpectralFlowwithBackreaction}
\end{figure}
\begin{figure}[t]
	\centering
		\includegraphics[width=0.8\linewidth]{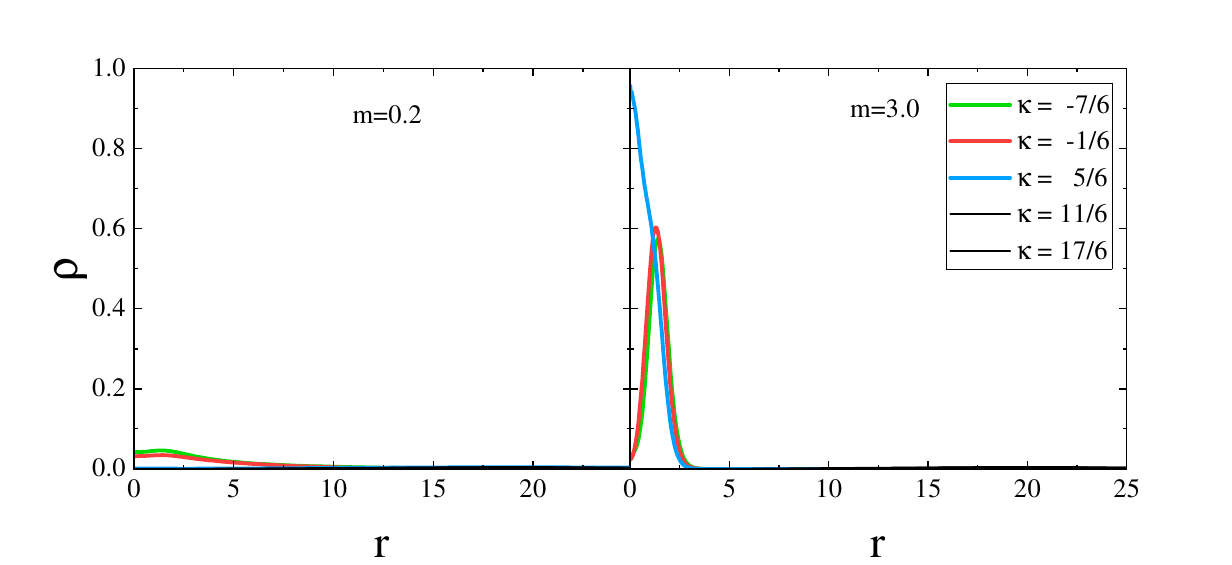}
		\caption{Fermion densities $\rho$ for each $\kappa$. \if0{The parameters are $\qty(M,e,\mu)=\qty(1.0,0.1,1.0)$, and the topological charge $Q=3$. }\fi Left panel : $m=0.2$. Right panel : $m=3.0$. Note that only $\kappa=\frac{5}{6}$ (blue curve) mode is coupled to the Skyrmion.}
		\label{FermionDensitywithbackreaction}
\end{figure}

In this section, we present our typical results. Hereafter, the parameters in the bosonic sector $(M,e,\mu)$ are fixed to $(M,e,\mu)=\qty(1.0,0.1,1.0)$. Since axisymmetric ansatz for Skyrmion is introduced in the previous section, the fermion field $\psi$ is also parametrized with axial symmetry. Recalling that the Skyrmions are static, we impose that the fermion field $\Psi$ has the stationaly form $\Psi=\psi e^{-i\epsilon t}$. Then, the Dirac equation \eqref{Fermioniceq} becomes 
\begin{align}
	\mathcal{H}\psi=\epsilon\psi,\ \mathcal{H}=\hat{\gamma}^0\qty(-i\hat{\gamma}^j\partial_j+m\hat{U}).\label{Fermioniceq2}
\end{align}
The corresponding eigenfunction of the Dirac Hamiltonian $\mathcal{H}$ can be written as 
\begin{align}
	\psi
	=\mathcal{N}\mqty
	(
		u_1\qty(r)e^{i\ell\varphi}\\
		d_1\qty(r)e^{i\qty(\ell+1)\varphi}\\
		s_1\qty(r)e^{i\qty(\ell+3)\varphi}\\
		u_2\qty(r)e^{i\qty(\ell+1)\varphi}\\
		d_2\qty(r)e^{i\qty(\ell+2)\varphi}\\
		s_2\qty(r)e^{i\qty(\ell+4)\varphi}
	)
	\label{Fermionicansatz}
\end{align}
where $u_{1,2}$, $d_{1,2}$ and $s_{1,2}$ are the radial functions of spinor components, $\ell\in\mathbb{Z}$ and $\mathcal{N}$ is the normalization factor which is defined in 
\begin{align}
	\int\dd[2]{x}\psi^\dagger\psi=2\pi\mathcal{N}^2\int_{0}^{\infty}\dd{r}r\qty(u_1^2+d_1^2+s_1^2+u_2^2+d_2^2+s_2^2)\equiv 1.
\end{align}

In this parametrization, the Dirac Hamiltonian $\mathcal{H}$ commutes with an extended spin operator $\mathcal{K}$, so called grand-spin operator, which is defined by 
\begin{align}
	\mathcal{K}=-i\partial_\varphi+\frac{1}{2}\sigma_3+\frac{1}{2}\lambda_3+\frac{5}{2\sqrt{3}}\lambda_8
\end{align}
where $\lambda_3$ and $\lambda_8$ represent the third and eighth components of Gell-Mann matrices respectively. Therefore, the solutions of eq.~\eqref{Fermioniceq2} are labeled by the eigenvalue $\kappa$ which is given by
\begin{align}
	\mathcal{K}\psi=\kappa\psi,\ \kappa=\ell+\frac{1}{2}+\frac{4}{3}.
\end{align}

One of the typical numerical results is presented in Fig.\ref{SpectralFlowwithBackreaction}, which shows the energy spectra for several choices of the coupling constant $m$. As one can see in Fig.\ref{SpectralFlowwithBackreaction}, there are three zero-crossing modes from negative to positive. This behavior is called spectral flow. These modes have corresponding quantum numbers $\kappa=\frac{5}{6},-\frac{1}{6},-\frac{7}{6}$ respectively. The number of zero-crossing modes equal to the topological charge $Q=3$ of the Skyrmion. This fact is a consequence of the Atiyah-Singer index theorem~\cite{Atiyah:1963zz}. Note that in order to input the fermion field into the fermion-Skyrmion interaction term of equation \eqref{Bosoniceq}, it is necessary to choose the grand spin $\kappa$. Here we consider the case of $\kappa=\frac{5}{6}\qty(\ell=-1)$.

\begin{figure}[t!]
	\begin{minipage}[b]{0.5\linewidth}
		\centering
		\includegraphics[width=1\linewidth]{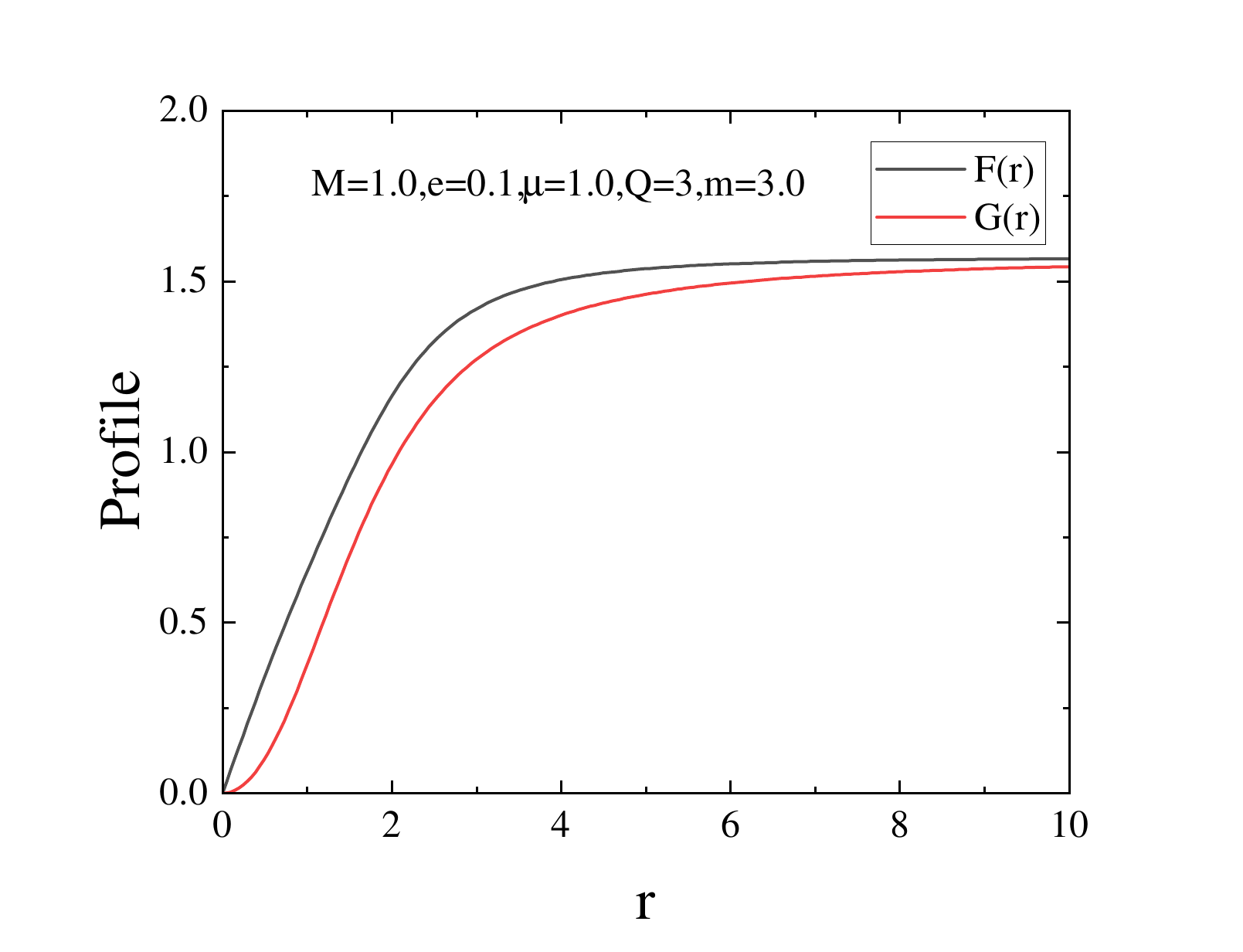}
	\end{minipage}
	\begin{minipage}[b]{0.5\linewidth}
		\centering
		\includegraphics[width=1\linewidth]{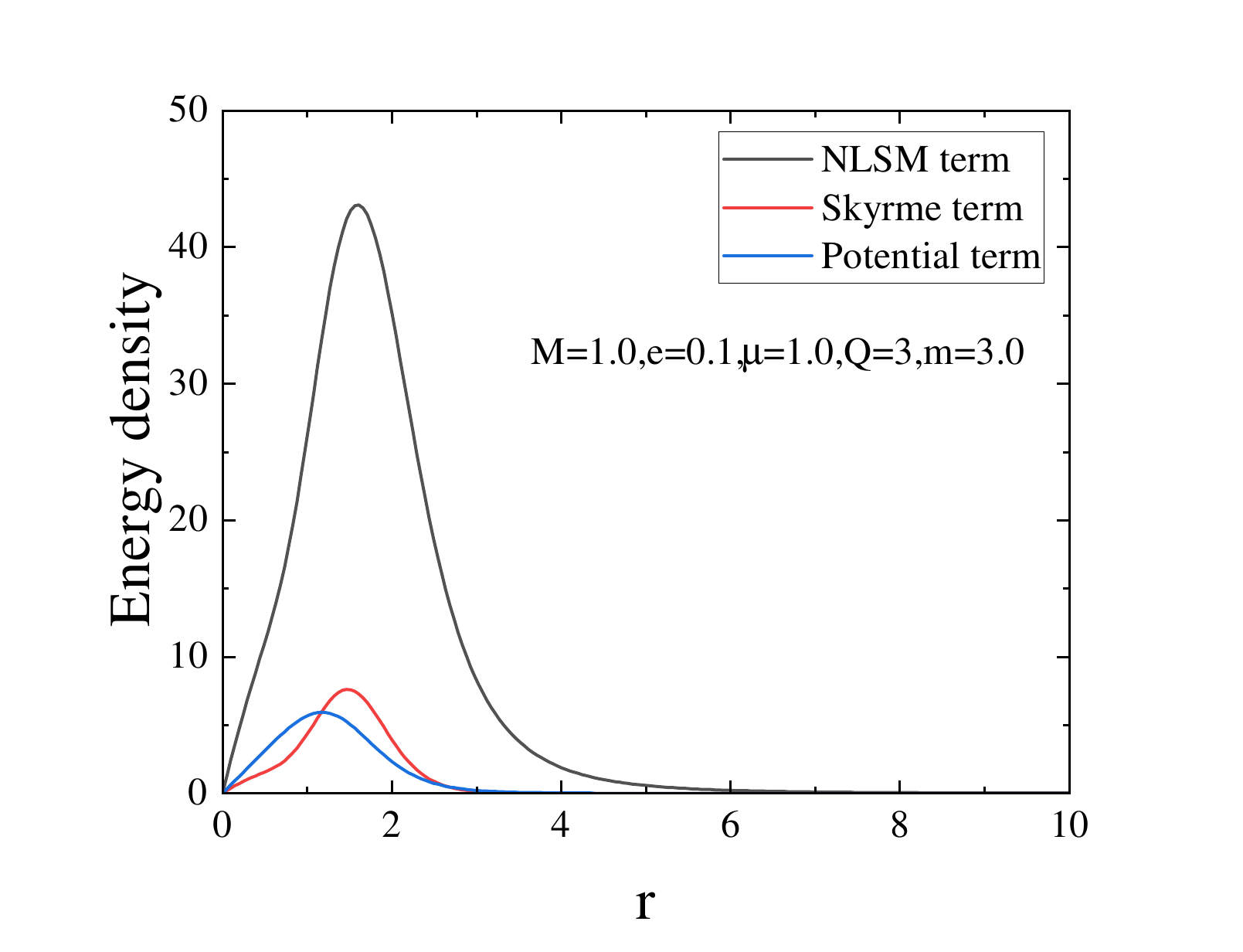}
	\end{minipage}
	\caption{Skyrmion solution with fermion backreaction. Left panel : Profile functions, $F$ and $G$. Right panel : Energy denitie of each terms in the bosonic energy, NLSM term, Skyrme term and potential term.}
	\label{BwithBackreactionm3}
\end{figure}

\begin{figure}[t!]
	\begin{minipage}[b]{0.5\linewidth}
		\centering
		\includegraphics[width=1\linewidth]{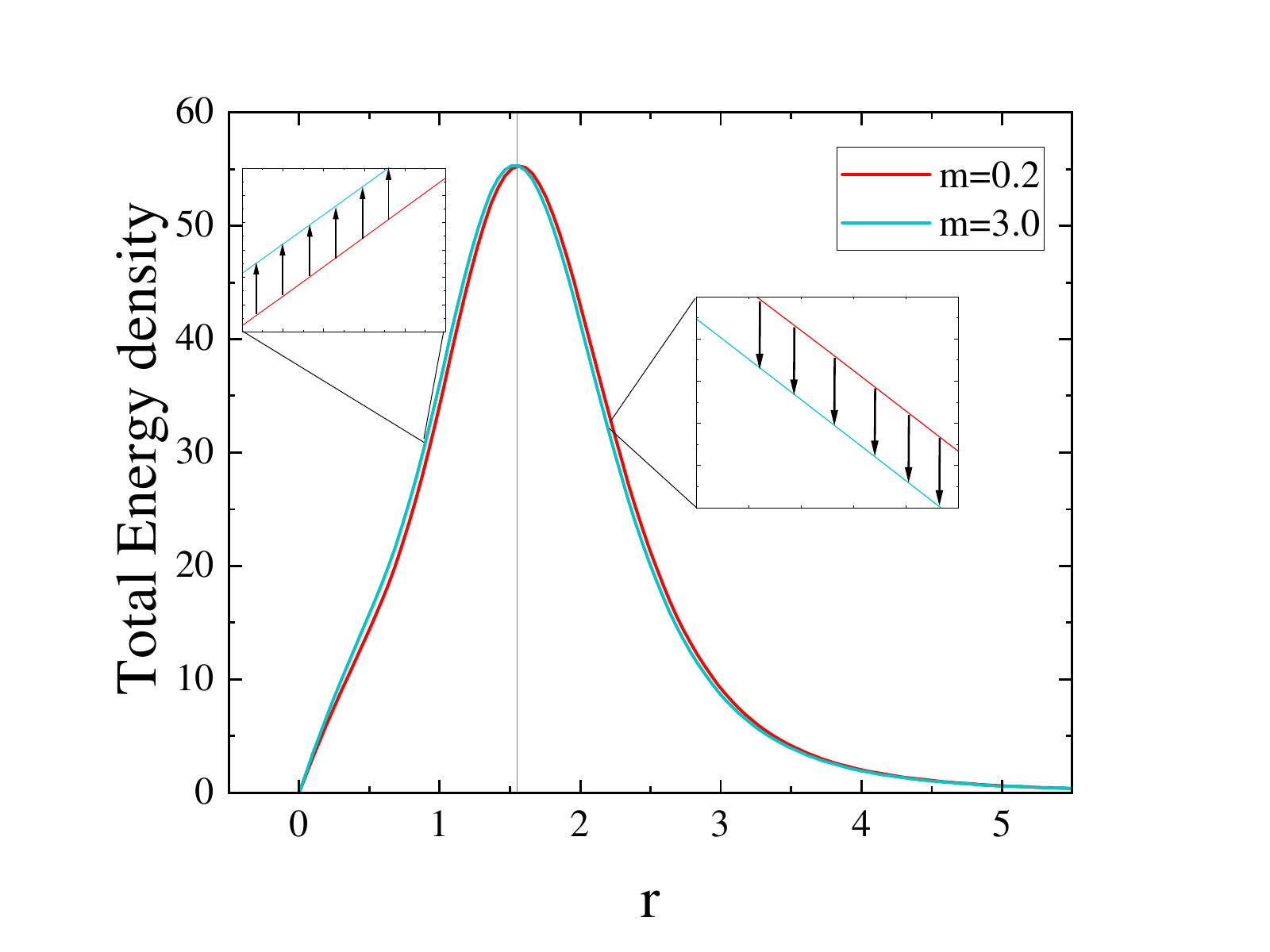}
	\end{minipage}
	\begin{minipage}[b]{0.5\linewidth}
		\centering
		\includegraphics[width=1\linewidth]{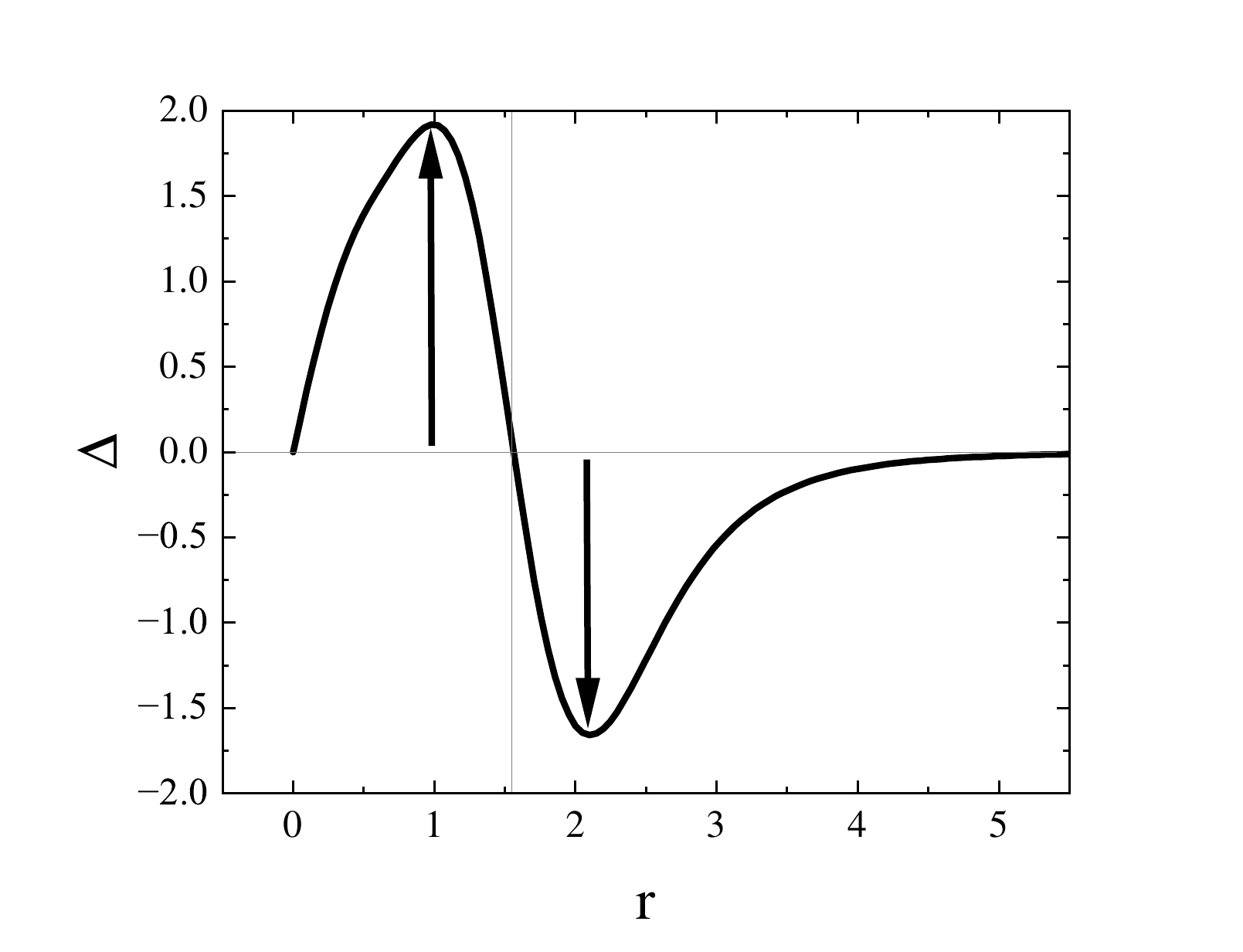}
	\end{minipage}
	\caption{Fermion backreaction effect. Left panel: Total energy densities $\mathcal{E}_b$ for $m=0.2$ and $m=3.0$. Right panel : $\Delta\equiv\mathcal{E}_b\vert_{m=3.0}-\mathcal{E}_b\vert_{m=0.2}.$}
	\label{BwithBackreaction}
\end{figure}

The zero-crossing modes are bound states. As shown in Fig.\ref{FermionDensitywithbackreaction}, the corresponding fermion density $\rho$ which is defined in 
\begin{align}
	\rho\qty(r)\equiv\frac{1}{2\pi}\int_0^{2\pi}\dd{\varphi}\psi^\dagger\psi
\end{align}
is localized around the Skyrmion center. Increasing the coupling constant $m$, the fermion density $\rho$ becomes strongly localized. In some previous studies, it is shown that this kind of localized fermions can deform the Skyrmion~\cite{Perapechka:2018yux,Dzhunushaliev:2024kti}. Such an effect is called backreaction effect. The bosonic field is shown in Fig.\ref{BwithBackreactionm3}. To see this, we compare the Skyrmion energy density $\mathcal{E}_b=-\mathcal{L}_b$ for weak and strong coupling constant $m$ in Fig.~\ref{BwithBackreaction}. The left panel shows the total bosonic energy density for $m=0.2,3.0$. The right panel plots the difference between the two densities $\Delta\equiv\mathcal{E}_b\vert_{m=3.0}-\mathcal{E}_b\vert_{m=0.2}.$ From this result, one sees that the backreaction makes the Skyrmion strongly localized. In other words, the Skyrmion-fermion coupling in our model \eqref{FullLagrangian} corresponds to an attractive force.


\section{Summary and Outlook}
\label{Summary}
In this paper, we construct some solutions of the $\CP^2$ Skyrmion with fermions. 
We construct fermion solutions on the Skyrmion. Then we observe the energy spectra have special behavior, so called spectral flow. There are some zero-crossing modes, and the number of such modes corresponds to the topological charge $Q$ of the Skyrmion. The zero-crossing modes are spatially localized around the Skyrmion. These localized modes can deform the Skyrmion through the backreaction effect. Then one observes that the Skyrmion deform increasing the fermion-Skyrmion coupling constant $m$. 
Particularly, Skyrmions are concentrated at the origin. This indicates that the backreaction corresponds to an attractive force. 

When fermions are absent, the Skyrmion solution is stabilized by the balance between Skyrme term and potential term. Our results show that the fermion-Skyrmion interaction has a similar effect to the potential term, so it may be possible to replace the role of the potential term 
with the backreaction. In fact, 
$\CP^1$ Skyrmions with localized fermions exist even in the absence of potential terms~\cite{Perapechka:2018yux}.  
We will analyze this possibility in the $\CP^2$ case and reconsider the Derrick's scaling argument~\cite{Derrick:1964} including isorotating fermions in the near future.

Furthermore, one can also consider stronger backreaction by introducing a coupling constant 
$g$ 
as $\mathcal{L}=\mathcal{L}_f+g^{-2}\mathcal{L}_b$. 
As the coupling constant $g$ increases, the energy scales of the fermions and the Skyrmion become closer, thereby enhancing the backreaction effect. 
Analysis of such strong backreaction effect is an important direction for future research.


\section*{Acknowledgments}
SY would like to thank all the conference organizers of ISQS29 and Prof.\v{C}estmir Burd\'{i}k for the hospitality and also many kind considerations. 
SY also thank the Yukawa Institute for Theoretical Physics at Kyoto University. 
Discussions during the YITP workshop ``Strings and fields 2025'' were useful to complete this work. The authors are also grateful Luiz Agostinho Ferreira, Pawel Klimas and Yakov Shnir for their many useful comments. 
This work is supported in part by JSPS KAKENHI [Grants No. JP23KJ1881 (YA), JP20K03278 (NS)] and JST SPRING [Grant Number JPMJSP2151 (SY)]. 


\section*{References}
\bibliography{backreaction}

\end{document}